# An Inter- and Intra-Band Loss for Pansharpening Convolutional Neural Networks

Jiajun CAI, Bo Huang

*Abstract*—Pansharpening aims to fuse panchromatic and multispectral images from the satellite to generate images with both high spatial and spectral resolution. With the successful applications of deep learning in the computer vision field, a lot of scholars have proposed many convolutional neural networks (CNNs) to solve the pansharpening task. These pansharpening networks focused on various distinctive structures of CNNs, and most of them are trained by L2 loss between fused images and simulated desired multispectral images. However, L2 loss is designed to directly minimize the difference of spectral information of each band, which does not consider the inter-band relations in the training process. In this letter, we propose a novel inter- and intra-band (IIB) loss to overcome the drawback of original L2 loss. Our proposed IIB loss can effectively preserve both inter- and intra-band relations and can be directly applied to different pansharpening CNNs.

## I. INTRODUCTION

Pansharpening is one of the fundamental techniques to improve the quality of remote sensing images. Due to the difficulties in obtaining satellite images with the both high spatial and spectral resolution, sensors equipped in the satellites will synchronously generate pairs of a low spatial resolution multispectral (MS) image and a high spatial resolution panchromatic (PAN) image captured in the same areas. Pansharpening is thus designed to generate pan-sharpened MS images that keep the same spatial resolution as PAN images.

A lot of pansharpening methods have been proposed in recent years. Most of these methods can be divided into three categories: component substitution-based, multiresolution analysis-based, and learning-based. The classical component substitution-based methods usually adopt the Brovey transform (BT) [1], the intensity-hue-saturation (IHS) [2], or principal component analysis (PCA) [3] to extract the main component of MS image and replace it by the PAN image to generate a pan-sharpened result. Multiresolution analysis-based methods use the various wavelet transforms [4]-[6] to decompose MS and PAN images into a series of sub-bands, and the fusion procedure is performed on these corresponding sub-bands from source images. The biggest advantage of aforementioned two categories is their computation efficiency, but they also easily render spectral distortion which will lose actual spectral information from original MS images.

Before deep learning is applied to pansharpening, variational optimization and dictionary learning are representative learning-based approaches that have been widely studied. Variational optimization-based methods involve a loss function and some prior regularization terms [7][8] to iteratively optimize fusion results. Dictionary learning-based methods [9][10] will firstly study dictionaries from training samples, and then replace original images by representation coefficients based on these dictionaries to perform fusion process.

With a lot of successful applications of deep learning, especially the convolutional neural network (CNN), in computer vision areas, many scholars also proposed various CNN architectures to deal with the pansharpening task. According to the similarities between pansharpening and super-resolution, pansharpening CNN (PNN) [11], which has a similar structure as super-resolution CNN (SRCNN) [12], is firstly proposed to bridge the deep learning and pansharpening. Combining domain knowledge, PanNet [13] is developed to perform the learning process in high-frequency bands using a ResNet-like structure [14]. Yuan et al. [15] incorporated the multi-scale and multi-depth idea into CNN and proposed MSDCNN. Following the development of deep learning, the architectures of networks become deeper and more complex.

Since most pansharpening CNNs still use L2 loss (Mean Squared Error) to minimize differences between fusion results and simulated ground truth MS images, it only calculates and optimizes the error between bands with the same wavelength. However, remote sensing images contain abundant spectral information, and adjacent bands are highly correlated. Obviously, these inter-band relations are not considered in current L2 loss.

In this letter, we propose a novel loss function based on the original L2 loss, named intra- and inter-band (IIB) loss. Our IIB loss includes two parts to regulate each band in the fused images: an intra-band loss which emphasizes keeping it the same as the corresponding band in the target MS image, and an inter-band loss which focuses on reconstructing the same inter-band relations as the target MS image. Fig. 1 adopts three bands images as an example to show an overall framework of a pansharpening CNN with our proposed IIB loss.

J. Cai and B. Huang are with the Department of Geography and Resource Management, The Chinese University of Hong Kong, Shatin, Hong Kong (e-mail: cai_jiajun@foxmail.com, bohuang@cuhk.edu.hk).

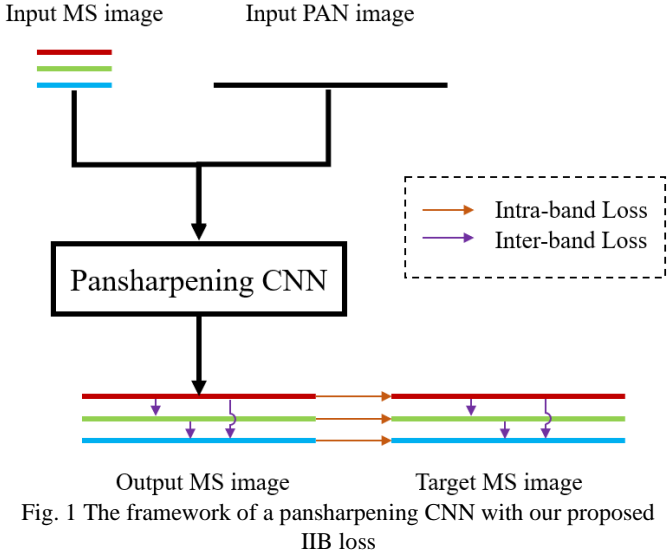

Fig. 1 The framework of a pansharpening CNN with our proposed IIB loss

### A. Proposed IIB Loss

Since the proposed IIB loss can be directly incorporated with existed pansharpening CNNs, we firstly define the universal CNN-based pansharpening process as

$$F = g(\downarrow M, \downarrow P; \theta) \quad (1)$$

Where $g$ is the pansharpening CNN which is parametrized by $\theta$. If L2 loss is adopted to optimize parameters in the network, the optimal $\theta$ is obtained by

$$\theta = \underset{\theta}{\arg\min} \sum_{i=1}^{I} \sum_{b=1}^{B} \left\| f_b^{(i)} - m_b^{(i)} \right\|_2^2 \quad (2)$$

where $f^{(i)} = g(\downarrow m_i, \downarrow p_i; \theta)$, and $(\downarrow m_i, \downarrow p_i, m_i)$ is the $i$th training sample. $B$ indicates the total number of bands.

Observing the form of L2 loss, we can find it will only calculate the differences between the same band within fusion results and target images. Prior and concurrent works [11]-[15] have proven the reliability of adopting L2 loss to optimize the whole network and then generate convincing fusion results. Therefore, we also use the original L2 loss to maintain intra-band relations in our designed IIB loss,

$$L_{Intra} = \sum_{i=1}^{I} \sum_{b=1}^{B} \left\| f_b^{(i)} - m_b^{(i)} \right\|_2^2 \quad (3)$$

For the inter-band relations, inspired by the QNR (quality with no reference) [17], we propose an inter-band loss which supports the training of pansharpening CNNs as follows:

$$L_{Inter} = \sum_{i=1}^{I} \sum_{l=1}^{B-1} \sum_{n=l+1}^{B} \left\| Q(f_l^{(i)}, f_n^{(i)}) - Q(m_l^{(i)}, m_n^{(i)}) \right\|_2^2 \quad (4)$$

where $Q$ is the universal image quality index [18] which is calculated by

$$Q(x,y) = \frac{4 \cdot \sigma_{xy} \cdot \bar{x} \cdot \bar{y}}{(\sigma_x^2 + \sigma_y^2)(\bar{x}^2 + \bar{y}^2)} = \frac{\sigma_{xy}}{\sigma_x \cdot \sigma_y} \times \frac{2 \cdot \bar{x} \cdot \bar{y}}{\bar{x}^2 + \bar{y}^2} \times \frac{2 \cdot \sigma_x \cdot \sigma_y}{\sigma_x^2 + \sigma_y^2} \quad (5)$$

in which $x$ and $y$ are images that need to be measured, and $\bar{x}$ and $\bar{y}$ are their corresponding means. $\sigma_x^2$ and $\sigma_y^2$ are the variance of $x$ and $y$, and $\sigma_{xy}$ denotes the covariance between $x$ and $y$. As shown in the right of equation (5), $Q$ can be decomposed into three factors. The first factor measures the correlation coefficient between $x$ and $y$, and it has a value range of [-1,1]. The second and third factors measure the luminance and contrast between $x$ and $y$, and they both have a value range of [0,1]. Therefore, $Q$ will equal to 1 if and only if $x=y$. In order to contain local statistics into consideration, $Q$ is calculated with a $W \times W$ sliding window, and the global score is averaged by these local values. Then, by combining $L_{Intra}$ and $L_{Inter}$, the proposed IIB loss can be written as

$$L_{IIB} = L_{Intra} + \alpha \cdot L_{Inter} \quad (6)$$

where $\alpha$ controls the importance of inter-band constraint, which is empirically set to 1.

## II. EXPERIMENTAL RESULTS AND DISCUSSION

### A. Datasets and Experimental Setting

We prepared two different datasets which include images from QuickBird (QB) and Worldview-3 (WV3), respectively. The spatial resolution of MS and PAN images from QB is 2.8m and 0.7m. The QB MS images include four bands: Infrared, Red, Green, and Blue. The spatial resolution of MS and PAN images from WV3 is 1.24m and 0.31m, while its MS bands covered by the wavelength of the panchromatic band are selected, which includes Infrared, Red, Yellow, Green, and Blue. Both QB and WV3 datasets consist of 7000 extracted training MS patches of size $64 \times 64$ and their corresponding PAN patches of size 256×256. We also prepared 200 MS patches of size 256×256 and their corresponding PAN patches of size 1024×1024 for testing.

The training datasets will be preprocessed according to Wald's protocol mentioned in Section II-A. The testing datasets can be organized in two forms. The first form consists of images prepared according to Wald's protocol, which is called simulated data. The second form directly uses original images, so it is named as actual data.

For simulated data, due to the existence of target images, we adopt indicators, including SAM [19], ERGAS [20], and UIQI [18], which need a full reference to evaluate the performance of different settings. Since there is no reference for actual data, QNR [17] with the spectral distortion index $D_\lambda$ and spatial distortion index $D_s$ are used for evaluating pansharpening results.

The effectiveness of our proposed IIB loss is proved by applying it to three representative pansharpening CNNs: PNN [11], DiCNN [21], and PanNet [13]. All deep learning-based methods are implemented on the GPU (NVIDIA GeForce RTX 2080Ti) through an open deep learning framework Tensorflow [22].

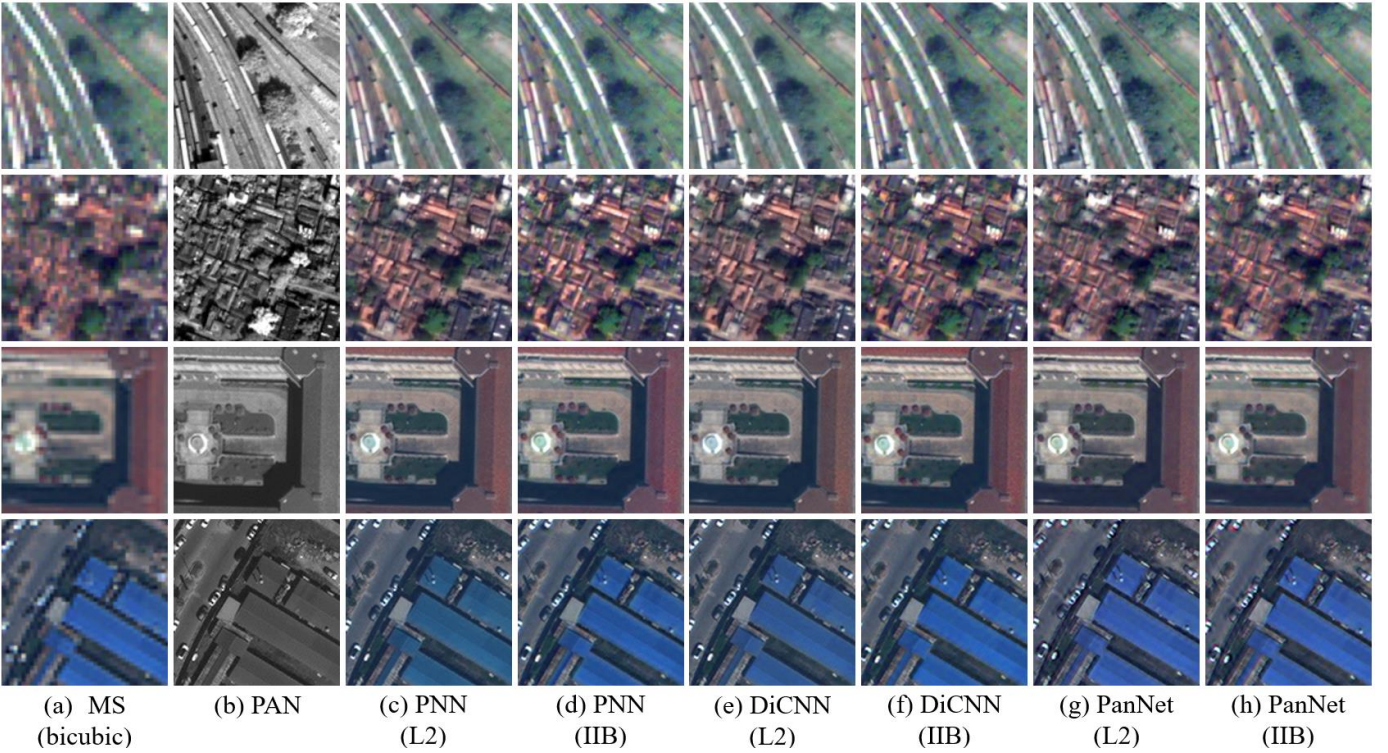

| | (a) MS (bicubic) | (b) PAN | (c) PNN (L2) | (d) PNN (IIB) | (e) DiCNN (L2) | (f) DiCNN (IIB) | (g) PanNet (L2) | (h) PanNet (IIB) |

Fig. 2 Fused results on the actual QB and WV3 data (zoomed in views).

TABLE I
EVALUATION OF FUSION RESULTS BASED ON TWO HUNDRED QUICKBIRD IMAGES

| | UIQI (↑) | SAM (↓) | ERGAS (↓) | $D_\lambda$ (↓) | $D_s$ (↓) | QNR (↑) |
|---|---|---|---|---|---|---|
| PNN | **0.7899** | **6.4564** | **5.9102** | 0.1045 | 0.0731 | 0.8299 |
| PNN+IIB | 0.7700 | 7.1522 | 6.3612 | **0.0442** | **0.0602** | **0.8982** |
| DiCNN | **0.7912** | **6.4351** | **5.8829** | 0.1013 | 0.0727 | 0.8332 |
| DiCNN+IIB | 0.7731 | 6.9219 | 6.2009 | **0.0397** | **0.0498** | **0.9124** |
| PanNet | **0.8012** | **6.4400** | **5.9498** | 0.0780 | 0.0457 | 0.8796 |
| PanNet+IIB | 0.7930 | 6.7040 | 6.0090 | **0.0326** | **0.0291** | **0.9392** |

TABLE II
EVALUATION OF FUSION RESULTS BASED ON TWO HUNDRED WORLDVIEW-3 IMAGES

| | UIQI (↑) | SAM (↓) | ERGAS (↓) | $D_\lambda$ (↓) | $D_s$ (↓) | QNR (↑) |
|---|---|---|---|---|---|---|
| PNN | **0.8157** | **5.7234** | **5.2639** | 0.0956 | 0.1428 | 0.7836 |
| PNN+IIB | 0.7918 | 6.7111 | 5.6357 | **0.0258** | **0.1074** | **0.8716** |
| DiCNN | **0.8170** | **5.6564** | **5.2391** | 0.0692 | 0.1208 | 0.8213 |
| DiCNN+IIB | 0.8022 | 6.2730 | 5.4791 | **0.0234** | **0.1028** | **0.8772** |
| PanNet | **0.8192** | **5.8299** | **5.2996** | 0.0606 | 0.0973 | 0.8498 |
| PanNet+IIB | 0.8048 | 6.1826 | 5.5236 | **0.0192** | **0.0683** | **0.9143** |

*B. Comparisons and Analysis*

In this subsection, we will apply our proposed IIB loss to different pansharpening CNNs and observe their corresponding performances.

Tables I and II summarize the objective evaluation based on QB and WV3 datasets at simulated and actual scales, where the up or down arrow indicates the higher or lower the better. We can notice that Tables I and II show a similar pattern. Observing original performances, PanNet obtains the best results in both simulated and actual scale. For the simulated datasets, the original L2 loss can obtain better UIQI, SAM, and ERGAS values since these indicators are averaged based on band-by-band results, but results generated by our IIB loss can still obtain close values. For the actual datasets, it can be found that the inter-band relations studied in the simulated scale have been successfully transferred to the actual scale. The values of $D_\lambda$, $D_s$, and QNR get dramatic improvement after applying the proposed IIB loss.

Fig. 2 shows the visual results of different settings, where the first and second rows are QB images, and the third and fourth rows are WV3 images. If we compare the pansharpening results generated by L2 loss and original MS images first, spectral preservation achieved by different networks is not satisfying enough. From the spatial perspective, the proposed IIB loss does not change spatial details comparing to results obtained by L2 loss, which means the intra-band restriction contained in IIB loss has the relatively same ability to generate pan-sharpened MS images. However, the spectral information is saliently adjusted when we apply IIB loss to pansharpening CNNs. The obvious spectral distortion can be observed in results generated by PNN and DiCNN when they are trained by L2 loss. The spectral residual module is widely adopted in the pansharpening CNNs, like DiCNN and PanNet, which are proposed after PNN to directly obtain spectral information from input LMS images. Although DiCNN shows better results than PNN, it still cannot avoid spectral distortion if we observe red and blue rooftops in the WV3 dataset. When the inter-band restriction is added to network training, we can find the spectral information is corrected even in the network without spectral residual module (PNN). This phenomenon highlights the importance of including the maintenance of inter-band relations in the spectral preservation strategy.

## III. Conclusion

In this letter, we propose an inter- and intra-band (IIB) loss for pansharpening CNNs. The biggest superiority of IIB loss is that it inherits the advantages of intra-band loss, e.g. L2 loss, and considers the inter-band restriction when we train a specific pansharpening CNN. Experimental results prove the effectiveness of preserving both intra- and inter-band relations by applying IIB loss.